\documentclass[prd, 10pt,aps,nofootinbib, floatfix, notitlepage,twocolumn,longbibliography,superscriptaddress]{revtex4-2}

\usepackage{amsmath,amsfonts,amssymb,enumitem}
\usepackage{mathrsfs}
\usepackage{graphicx}
\usepackage[english]{babel} 
\usepackage{url} 
\usepackage{bbold}
\usepackage[utf8]{inputenc} 
\usepackage[colorlinks=true,citecolor=blue,urlcolor=blue]{hyperref}
\usepackage{bm}% bold math
\usepackage[dvipsnames]{xcolor}
\usepackage{multirow} 
\usepackage{float}
\usepackage{mathrsfs}

\hypersetup{linktocpage=true }

\newcommand{\LCDM}{$\Lambda$CDM}

\newcommand{\Col}{@{\hspace{0.6em}} c @{\hspace{0.6em}}}
\newcommand{\sfrac}[2]{\dfrac{\,#1\,}{\,#2\,}} 

\newcommand{\mpl}{M_{\rm Pl}}

\begin{document}

\title{
An \textit{Attractive} Proposal for Resolving the Hubble Tension:~Dynamical Attractors that Unify Early and Late Dark Energy}
\author{Omar F. Ramadan}
\email{oramadan@hawaii.edu}
\affiliation{%
Department of Physics $\&$ Astronomy, University of Hawai‘i,
Watanabe Hall, 2505 Correa Road, Honolulu, HI, 96822, USA}%
\author{Tanvi Karwal}
\email{karwal@sas.upenn.edu}
\affiliation{%
Center for Particle Cosmology, Department of Physics and Astronomy,
University of Pennsylvania, Philadelphia, PA 19104, USA}%
\author{Jeremy Sakstein}
\email{sakstein@hawaii.edu}
\affiliation{%
Department of Physics $\&$ Astronomy, University of Hawai‘i,
Watanabe Hall, 2505 Correa Road, Honolulu, HI, 96822, USA}%
\date{\today}

\begin{abstract}
Early dark energy is a promising potential resolution of the Hubble tension.~Unfortunately, many models suffer from the need to fine-tune their initial conditions to ensure that the epoch of early dark energy coincides with matter-radiation equality.~We propose a class of \textit{attractive early dark energy} models where this coincidence arises naturally as a saddle point of a dynamical system that attracts a large volume of phase-space trajectories regardless of the initial conditions.~The system approaches a global dark energy attractor at late-times.~Our framework therefore unifies early and late dark energy using a single scalar degree of freedom.~We analyze a fiducial attractive early dark energy model and find that it is disfavored by cosmological data due to the presence of a long-lived saddle point in the matter era where the scalar plays the role of an additional component of (non-clustering) dark matter.~Our investigations provide lessons for future model-building efforts aimed at constructing viable attractive early dark energy models.~
\end{abstract}
\maketitle

\section{Introduction}\label{Section I:Intro}

Discovering the origin of the Hubble tension --- which refers to the statistically significant discrepancy between measurements of the Hubble constant $H_0$ made using late-universe probes and early-universe inferences \cite{Verde:2019ivm, Kamionkowski:2022pkx, Poulin:2023lkg} --- is an urgent goal of cosmology research.~The lack of a complete concordance model that is capable of accommodating all of our observations is limiting our ability to interpret data from cosmological missions and astrophysical surveys, and will continue to do so until the physics responsible for the Hubble tension is identified.

The strongest tension, now over $5\sigma$, is between the  $\Lambda$CDM fit to the Planck cosmic microwave background (CMB) data, which yields $H_0=\left(67.4\pm0.5\right)$km/s/Mpc \cite{Planck:2018vyg};~and the measurement by the SH0ES collaboration, who report  $H_0=\left(73.29\pm0.90\right)$km/s/Mpc \cite{Murakami:2023xuy} using a Cepheid and type-Ia supernova distance ladder.~While there is some spread in the late-universe measurements of $H_0$, they  trend to higher $H_0$ values, with none scattering lower than the Planck CMB estimate \cite{Abdalla:2022yfr}.~Similarly, inferences based on the early universe, even those independent of any CMB data, are clustered at low $H_0$ values \cite{Schoneberg:2019wmt,eBOSS:2020yzd,DAmico:2019fhj,Philcox:2020vvt}, 
with none scattering higher than the SH0ES late-universe measurement.~
Concerted efforts to update and interrogate the data over the past few years have failed to find any evidence that the tension is due to a systematic error in  the data \cite{Riess:2021jrx,Yuan:2019npk,Soltis:2020gpl,Anand:2021sum,Anderson:2023aga,Shajib:2023uig,Kenworthy:2022jdh}.~It is difficult to imagine a single systematic that would affect the various different objects used to calibrate the distance ladder, as each is governed by different physics.
Alternatively, a series of uncorrelated systematics must miraculously conspire to bias $H_0$ upwards by similar amounts in order to explain the tension.~Given these considerations, the hypothesis that the disagreement between early- and late-universe measurements of $H_0$ originates from new physics beyond the cosmological standard model has been the subject of an intense research effort \cite{DiValentino:2021izs,Poulin:2023lkg}.

A plethora of theoretical solutions have been proposed \cite{DiValentino:2021izs,Poulin:2023lkg,Schoneberg:2021qvd}.~Among them, early dark energy (EDE) \cite{Karwal:2016vyq,Poulin:2018dzj,Poulin:2018cxd} is one of the most successful \cite{Schoneberg:2021qvd}.~In this scenario, a new component of the Universe becomes active around the time of matter-radiation equality and accounts for a maximal fractional contribution $f_{\rm ede}\sim10\%$ of the Universe's energy budget at a critical redshift $z_c\sim 3300$.~The presence of this additional component increases the pre-recombination expansion rate, which shrinks the \textit{physical} size of the sound horizon.~To compensate and maintain the precisely measured \textit{angular} size $\theta_*$ of the sound horizon, the CMB-predicted $H_0$ increases toward the locally-measured value.~Post recombination, the EDE must redshift away faster than radiation in order to preserve the excellent \LCDM\ fit to late-time cosmic observables \cite{Poulin:2018cxd}.

Despite its success, EDE faces several challenges.~First, another growing tension in cosmology, the $\sim3\sigma$ $S_8$ tension between the amplitude of density fluctuations inferred from the CMB and measured by weak-lensing probes \cite{KiDS:2020suj}, is exacerbated by EDE  \cite{Hill:2020osr, Smith:2020rxx} because it predicts a universe with more dark matter than \LCDM.~Second, the physics of EDE is disconnected from the physics of photons and baryons, presenting a coincidence or \textit{why then?} problem.~Why was EDE important at matter-radiation equality and not some other epoch?~The majority of models achieve this coincidence by fine-tuning the model parameters but these are not protected from radiative corrections by any fundamental symmetries.~These models are therefore unnatural effective field theories.~Proposals that avoid these fine-tunings include models where the EDE scalar couples to neutrinos \cite{Sakstein:2019fmf,CarrilloGonzalez:2020oac,CarrilloGonzalez:2023lma} that naturally begin the epoch of EDE when they inject energy into the scalar when they become non-relativistic, which, coincidentally, is around the time of matter-radiation equality;~realizing EDE potentials in UV-complete theories such as string theory \cite{McDonough:2022pku, Andriot:2023wvg};~coupling EDE to dark matter \cite{Karwal:2021vpk,Lin:2022phm,McDonough:2021pdg};~and coupling it to a non-Abelian gauge group \cite{Berghaus:2022cwf,Berghaus:2019cls}.~Finally, the \textit{ad hoc} nature of EDE is  unappealing.~EDE is posited solely to solve the Hubble tension, and does not reach beyond that issue to connect with other cosmological phenomena such as late dark energy (LDE).~This final problem has led to attempts to unify EDE and LDE through quintessence models where a scalar field plays the role of both EDE and LDE \cite{Adil:2022hkj,Brissenden:2023yko, Chowdhury:2023opo}.

In this work, we propose a  model that attempts to address the shortcomings described above:~\textit{attractive Early Dark Energy} (@EDE).~
In this framework, both EDE and LDE are described by a single  quintessence scalar field $\phi$ with a non-linear potential that results in autonomous cosmological equations that form a  dynamical system.~The potential is chosen such that:~(1) EDE arises as a saddle point during the radiation epoch so that solutions naturally flow towards it regardless of the initial conditions;~and (2) the late-time global attractor is a dark-energy-dominated universe.~This framework overcomes the need for fine-tuning as the fixed points determine the magnitude of the EDE injection, not the initial conditions.~In addition, the framework is  appealing 
because unifying EDE and LDE reduces the number of extra 
degrees of freedom needed to explain the $H_0$ tension and late dark energy.~

We study an example of @EDE that is motivated by string theory, analyze its background dynamics, and confront it with cosmological data.~Ultimately, we find that this specific model is not preferred by the data due to its percent-level contribution to the energy budget of the post-recombination universe.~Our analysis reveals important lessons that lay the foundations for constructing viable @EDE models that we discuss in our conclusions.~

This paper is organized as follows:~In section (\ref{Section II.A:The model}), we lay out our  framework with a discussion of the dynamical system, its fixed points, and the properties they must possess in order to potentially resolve the $H_0$ tension and to drive LDE.~In section (\ref{Section II.B:Background dynamics}), we discuss the background dynamics of the field.~We outline our analysis methodology, describing our choices of parameters, priors and data sets in section (\ref{Section III:Methodology}) and present its results in section (\ref{Section IV:Results}).~Finally, in section (\ref{Section V:Discussion}), we discuss our results and draw conclusions.~

\section{Attractive Early Dark Energy}\label{Section II:@EDE}

\subsection{Framework and Model}\label{Section II.A:The model}

Our framework for constructing @EDE models is the dynamical systems formulation of a single uncoupled quintessence scalar $\phi$ \cite{Copeland:1997et,Copeland:2006wr,Bahamonde:2017ize}.~This formalism has been successful at alleviating the LDE coincidence problem, and provides a convenient and well-studied starting point for constructing dynamical systems that include EDE fixed points.~The action consists of the Einstein-Hilbert action, an uncoupled quintessence scalar field $\phi$, and a decoupled Standard Model (SM) and dark matter sector described by $S_{\rm SM}$ and $S_{\rm DM}$ respectively:~
\begin{eqnarray}
    S=&&\int d^4x\sqrt{-g}\left[\frac{\mpl^2R}{2}-\frac{1}{2}\nabla_\mu\phi\nabla^\mu\phi-V(\phi)\right]\nonumber\\ 
    +&&S_{\rm SM}+S_{\rm DM}.
\end{eqnarray}
We assume a flat Friedmann-Lema\^{i}tre-Robertson-Walker (FLRW) universe that is described by the metric 
\begin{equation}
	ds^2=-dt^2+a^2(t)\delta_{ij}dx^idx^j~\label{eq:metric},
\end{equation}
where $a(t)$ is the dimensionless scale factor normalized to unity today.~
The evolution of the scalar field is determined by the Klein-Gordon equation
\begin{equation}
    \Ddot{\phi}+3H\Dot{\phi}+\frac{dV}{d\phi}=0 \,.
    \label{eq:klein-gordon}
\end{equation}
We consider a universe containing matter $(m)$, radiation $(r)$, and the scalar;~and define the following quantities:
\begin{gather}
    \Omega_m=\frac{\rho_m}{3\mpl^2H^2},\label{eq:Omega_i}\quad \Omega_r=\frac{\rho_r}{3\mpl^2H^2},\\ \Omega_k=\frac{\dot{\phi}^2}{6\mpl^2H^2},\quad \,\,\,\,\,\Omega_v=\frac{V(\phi)}{3\mpl^2H^2} ,\label{eq:Omega_kv}\\
    w_\phi=\frac{P_\phi}{\rho_\phi}=\frac{\Omega_k-\Omega_v}{\Omega_k+\Omega_k},\\
    \lambda\equiv -\mpl\frac{V_{,\phi}}{V(\phi)},\label{eq:lambda}\\
    \Gamma\equiv \frac{V(\phi)V_{,\phi\phi}}{V_{,\phi}^2} \,.
    \label{eq:gammma}
\end{gather}
Here, $\Omega_m$, $\Omega_r$,
$\Omega_k$, and $\Omega_v$ are the density parameters for for matter, radiation, the  scalar's kinetic energy, and the scalar's potential energy respectively;~$w_\phi$ is the scalar's (time-dependent) equation of state;~and $\lambda(\phi)$ and $\Gamma(\phi)$ are the \textit{roll} and \textit{tracker} parameters respectively, which are helpful for parameterizing the dynamical system.~

\def\arraystretch{1.5}
\begin{table*}[t]    
    \begin{tabular}{|\Col|\Col|\Col|\Col|\Col|\Col|\Col|}\hline
			\# &  \boldmath $\Omega_\phi$ & \boldmath $\Omega_X$ & \textbf{Existence} & \textbf{Condition} & \textbf{Stability} & \boldmath $w_\phi$ \\ \hline\hline
                O$_{\lambda}$ & $0$ & $0$ & Always & None & Saddle & $0$ \\
                \hline
			\multirow{3}{*}{A$^*$}  & \multirow{3}{*}{$1$} & \multirow{3}{*}{$0$} & \multirow{3}{*}{$\forall \lambda_*$}  & $\lambda_* > -\sqrt{6}$ and $\Gamma_*'>0$ & Unstable node & \multirow{3}{*}{$1$}  \\
			&  & & & $\lambda_* < -\sqrt{6}$& Saddle & \\
                &  & & & $\Gamma_*'<0$ & Saddle & \\
			\hline
   
			\multirow{2}{*}{B$^*$} &  \multirow{2}{*}{$\sfrac{3(1+w)}{\lambda_*^2}$} & \multirow{2}{*}{$1-\sfrac{3(1+w)}{\lambda_*^2}$} & \multirow{2}{*}{$\lambda_* \geq \sqrt{3(1+w)}$} & $\lambda_*\Gamma_*'>0$ & Stable node & \multirow{2}{*}{$w$} \\
			& & & &  $\lambda_*\Gamma_*'<0$ & Saddle &  \\ 
			\hline
			\multirow{3}{*}{C$^*$}  & \multirow{3}{*}{$1$} & \multirow{3}{*}{$0$} & \multirow{3}{*}{$1$} & $0$ and $\lambda_*\Gamma_*'>0$ & Stable node& \multirow{3}{*}{$\!-\! 1+\sfrac{\lambda_*^2}{3}$} \\
			& & & & $\sqrt{3(1+w)}\leq\lambda_* < \sqrt{6}$& Saddle & \\
            & & & & $\lambda_*\Gamma_*<0$& Saddle & \\
            \hline
            \multirow{2}{*}{D} &  \multirow{2}{*}{$1$} & \multirow{2}{*}{$0$} & \multirow{2}{*}{Always} & $\lambda^2\left[\Gamma(\lambda)-1\right]|_{\lambda=0}>0$ & Stable node & \multirow{2}{*}{$-1$} \\
			& & & &  $\lambda^2\left[\Gamma(\lambda)-1\right]|_{\lambda=0}<0$ & Saddle &  \\ 
			\hline
    \end{tabular}
    \caption{Fixed points of the general dynamical system when $\lambda(\phi)$ is invertible.~$\Omega_X$ represents the density parameter of the dominant species --- either matter or radiation --- i.e., $X=m,\,r$.~Each distinct value of $\lambda_*$ that solves equation \eqref{eq:attractor_roots} gives rise to a set of fixed points A*, B*, and C*.~$\Gamma_*'=\Gamma'(\lambda_*)$.~Note that point O$_\lambda$ exists independently of the potential, and point D has $\lambda=0$ so corresponds to a cosmological constant.~Point D always exists provided $\lambda=0$ is accessible.}
\label{tab:general_fixed_points} 
\end{table*}

The equations \eqref{eq:Omega_i}-\eqref{eq:gammma} along with the Friedmann equations and the continuity equations for matter and radiation can be written as a dynamical system with phase space $\{\Omega_m, \Omega_r, \Omega_k, \Omega_v, \lambda, \Gamma\}$ \cite{Copeland:2006wr,Bahamonde:2017ize}.~The Friedmann constraint, $\Omega_m+\Omega_r+\Omega_k+\Omega_v=1$, allows us to eliminate one of the $\Omega$'s in terms of the others, reducing the dimension of the phase space by one.~We choose to eliminate $\Omega_m$.~The equations of motion in dynamical systems form are then:
\begin{align} 
\Omega_r'&=\Omega_r(3\Omega_m+4\Omega_r+6\Omega_k-4)\label{eq:Omega_r'},\\
    \Omega_k'&=\Omega_k(3\Omega_m+4\Omega_r+6\Omega_k-6)+\lambda\Omega_v\sqrt{6\Omega_k}\label{eq:Omega_k'},\\
    \Omega_v'&=\Omega_v(3\Omega_m+4\Omega_r+6\Omega_k-\lambda\sqrt{6\Omega_k})\nonumber\\
    &+\lambda\Omega_v\sqrt{6\Omega_k}\label{eq:Omega_v'},\\
    \lambda'&=-\lambda^2(\Gamma(\phi)-1)\sqrt{6\Omega_k},\label{eq:lambda'} \,,
\end{align}
where $\Omega_m=1-\Omega_r-\Omega_k-\Omega_v$.~These equations do not form an autonomous system unless  either $\lambda'=0$ identically, $\lambda(\phi)$ is invertible so that one can find $\phi(\lambda)$  in closed form by inverting equation \eqref{eq:lambda} and  write $\Gamma(\phi)=\Gamma(\phi(\lambda))=\Gamma(\lambda)$ to close the system, or further equations for derivatives of $\Gamma$ are supplied.~The first case implies that $\lambda$ is constant, which corresponds to an exponential potential $V(\phi)=V_0\exp(-\lambda\phi/\mpl)$.~The resulting three-dimensional phase space of this system $\{\Omega_r, \Omega_k, \Omega_v\}$ has been extensively studied \cite{Copeland:1997et,Copeland:2006wr}, and it is not possible to simultaneously have an LDE attractor with $w_\phi \approx -1$ and an early-universe saddle point that could play the role of EDE.~The case where $\lambda(\phi)$ is invertible so that $\Gamma=\Gamma(\lambda)$ corresponds to a four-dimensional phase space $\{\Omega_r,\Omega_k,\Omega_v,\lambda\}$.~We will examine the structure of this phase space shortly and find that this dynamical system does admit the possibility of @EDE models with an EDE saddle and an LDE attractor.~Systems where derivatives of $\Gamma$ (and possibly their derivatives) are required correspond to higher-dimensional phase spaces.~We will not explore these systems here, but follow-up investigations along these lines would certainly be interesting.

\def\arraystretch{1.4}
\begin{table*}[t]
\begin{tabular}{|\Col|\Col|\Col|\Col|\Col|\Col|\Col|\Col|}\hline
			\# & \boldmath $\Omega_m$ & \boldmath $\Omega_r$ & \boldmath $\Omega_\phi$ & \textbf{Existence} & \textbf{Condition} & \textbf{Stability} & \boldmath $w_\phi$ \\ \hline\hline
			
			A & $0$ & $0$ & $1$ & $\forall \, \lambda_*$ & - & Saddle & $-1$ \\ 
			\hline
			\multirow{2}{*}{B} & \multirow{2}{*}{$0$} & \multirow{2}{*}{$0$} & \multirow{2}{*}{$1$} & \multirow{2}{*}{$\forall \, \lambda_*$} & $\lambda_* < \sqrt{6}$ & Unstable node & \multirow{2}{*}{$1$} \\ 
			& & & & & $\lambda_* > \sqrt{6}$ & Saddle &  \\ 
			\hline
			C & $0$ & $1$ & $0$ & $\forall \, \lambda_*$ & - & Saddle & $0$ \\ 
			\hline
			D & $1$ & $0$ & $0$ & $\forall \, \lambda_*$ & - & Saddle & $0$ \\ 
			\hline
			\multirow{3}{*}{E} & \multirow{3}{*}{$0$} & \multirow{3}{*}{$0$} & \multirow{3}{*}{$1$} & \multirow{3}{*}{$\lambda_* < \sqrt{6}$} & $0 < \lambda_* < \sqrt{3}$ & Stable node & \multirow{3}{*}{$\!-\! 1+\sfrac{\lambda_*^2}{3} $}  \\
			& & & & & $\sqrt{3} < \lambda_* < 2$ & Saddle & \\
			& & & & & $2 < \lambda_* < \sqrt{6}$ & Saddle & \\ 
			\hline
			\multirow{2}{*}{F} & \multirow{2}{*}{$1 \!-\! \sfrac{3}{\lambda_*^2}$} & \multirow{2}{*}{$0$} & \multirow{2}{*}{$\sfrac{3}{ \lambda_*^2}$} & \multirow{2}{*}{$\lambda_* > \sqrt{3}$} & $\sqrt{3} < \lambda_* < \sqrt{24/7}$ & Stable node & \multirow{2}{*}{$0$} \\
			& & & & & $\lambda_* > \sqrt{24/7}$ & Stable spiral &  \\ 
			\hline
			\multirow{2}{*}{G} & \multirow{2}{*}{$0$} & \multirow{2}{*}{$1 \!-\! \sfrac{4}{\lambda_*^2}$} & \multirow{2}{*}{$\sfrac{4}{\lambda_*^2}$} & \multirow{2}{*}{$\lambda_* > 2$} & $2 < \lambda_* < \sqrt{64/15}$ & Saddle & \multirow{2}{*}{$1/3$} \\
			& & & & & $\lambda_* > \sqrt{64/15}$ & Saddle spiral & \\
			\hline
    \end{tabular}
    \caption{Fixed points and the stability of the phase space of a single exponential quintessence potential.
    The points relevant to our model are the EDE saddle point G, the matter-domination stable spiral F, and the scalar field-dominated LDE attrator E.~Note that points F and G correspond to point B* in Table~\ref{tab:general_fixed_points} with $w$ fixed to the dominant species.} 
    \label{tab:fixed_points_table} 
    
\end{table*}
When $\lambda$ is invertible, the fixed points of the dynamical system 
corresponds to points where equations \eqref{eq:Omega_r'}-\eqref{eq:lambda'} are simultaneously equal to zero.~We can characterize these independently of the choice of potential  by setting equation \eqref{eq:lambda'} to zero and finding all roots $\lambda_*$ such that
\begin{equation}
    \lambda_*^2 (\Gamma\left(\lambda_*\right)-1)=0\,,
    \label{eq:attractor_roots}
\end{equation}
and substituting the roots $\lambda=\lambda_*$ into \eqref{eq:Omega_r'}-\eqref{eq:Omega_v'} to obtain the fixed points.~These are listed in Table~\ref{tab:general_fixed_points} along with their linear stability;~we refer the reader to \cite{Copeland:2006wr,Bahamonde:2017ize} for details of how the stability is determined.~

Examining Table~\ref{tab:general_fixed_points} reveals the properties of the potential required to construct @EDE models.~Specifically, we require that:
\begin{enumerate}
    \item Fixed point B* exists, and corresponds to an EDE saddle point i.e., $\lambda_*$ is such that $\Omega_\phi\sim10\%$.
    \item Fixed point C* exists and is a stable late-time global attractor that will account for LDE i.e., $\lambda_*$ is such that $w_\phi\approx-1$.\footnote{One could replace fixed point C* by fixed point D, but this corresponds to the scalar behaving as a pure cosmological constant i.e., late dark energy is non-dynamical.~This scenario is equivalent to the canonical EDE scenario.~}
\end{enumerate}
The above requirements are incompatible with a single value of $\lambda_*$ because having $w_\phi\approx-1$ at point C* implies a small $\lambda_*$ that would render point B* non-existent, so any viable potential must admit multiple distinct roots  of equation \eqref{eq:attractor_roots}.~

To make further progress, we must specify a potential, calculate $\Gamma(\lambda)$, and hence calculate $\lambda_*$ as a function of the potential's parameters.~ Only specific potentials give rise to functional forms for $\lambda(\phi)$ that can be inverted.~These include the single exponential ($\Gamma=1$) potential discussed above, power-law potentials, and the particular functions listed in Table $10$ of \cite{Bahamonde:2017ize}.~We surveyed the potentials known to have invertible $\lambda(\phi)$ and identified two that satisfy the criteria above:
\begin{align}
    V(\phi)&=V_\alpha e^{-\alpha\frac{\phi}{\mpl}}+V_\beta e^{-\beta\frac{\phi}{\mpl}}\label{eq:double_exp},\\
    V(\phi)&=V_0 \left(\eta+e^{-\alpha\frac{\phi}{\mpl}}\right)^{-\beta}.\label{eq:exp+cst}
\end{align}
The first is the double exponential \eqref{eq:double_exp} that has been thoroughly studied in the context of quintessence LDE \cite{Sen:2001xu,Jarv:2004uk,Li:2005ay} and  arises naturally in string theory \cite{Barreiro:1999zs}.~The second potential \eqref{eq:exp+cst} is contrived to correspond to a non-dynamical cosmological constant at late times \cite{Zhou:2007xp,Fang:2008fw}.~Wishing to focus on natural and well-motivated models, we  confine our study to the double exponential potential.~This potential shares some similarities with the \textit{assisted quintessence} EDE model studied by \cite{Sabla:2021nfy} in which multiple scalar fields, each with a single exponential potential, are investigated as a resolution of the Hubble tension.~

\subsection{Background dynamics}
\label{Section II.B:Background dynamics}

The double exponential potential has 
\begin{align}
    \lambda(\phi)=\frac{\alpha  V_{\alpha } e^{-\frac{\alpha  \phi }{M_{\rm Pl}}}+\beta  V_{\beta } e^{-\frac{\beta  \phi }{M_{\rm Pl}}}}{V_{\alpha } e^{-\frac{\alpha  \phi }{M_{\rm Pl}}}+V_{\beta } e^{-\frac{\beta  \phi }{M_{\rm Pl}}}}\label{eq:lambdaDoubleExponential}.
\end{align}
 Taking the limits $\phi\rightarrow\pm\infty$, it follows that $\lambda$ is bounded to lie in the interval $\beta\le\lambda\le\alpha$.~Inverting equation \eqref{eq:lambdaDoubleExponential}, one finds
\begin{equation}
    \label{eq:GammaDoubeExponential}
    \Gamma(\lambda)=1-\frac{(\alpha-\lambda)(\alpha-\beta)}{\lambda^2},
\end{equation}
implying that equation \eqref{eq:attractor_roots} has two roots:~$\lambda_*=\alpha$ and $\lambda_*=\beta$.~We make the arbitrary choice to impose $\alpha>\beta$, in which case we can identify the EDE saddle point B* with $\lambda_*=\alpha$ and the LDE attractor C* with $\lambda_*=\beta$.~We can gain insight into the cosmology of this potential by considering the limit  $V_{\alpha} \gg V_{\beta}$, motivated by the hierarchy between the energy scales of EDE and LDE.~We then expect that this region of parameter space corresponds to EDE driven by the $\alpha$-exponential and LDE driven by the $\beta$-exponential.~In this limit, the potential can be approximated as 
\begin{equation}
   V(\phi) \approx \begin{cases}
    V_\alpha e^{-\alpha\frac{\phi}{\mpl}}, & \text{at early times},\\
    V_\beta e^{-\beta\frac{\phi}{\mpl}}, & \text{at late times}, 
\end{cases}
\end{equation}
assuming appropriate initial conditions.~To a good approximation, this implies that
 \begin{equation}
   \lambda_* \approx \begin{cases}
    \alpha, & \text{at early times},\\
    \beta, & \text{at late times}.~
\end{cases}
\end{equation}
This simplifies the analysis because the $\lambda_*=\alpha$ saddle --- which corresponds to EDE  and exists whenever $\alpha>2$ --- dictates the cosmology of the early universe, and we shall hence refer to it as it the \textit{EDE saddle};~and the $\lambda_*=\beta$ attractor --- which corresponds to LDE  and exists whenever $\beta<\sqrt{6}$  --- dictates the late universe cosmology, so we refer to this as the \textit{LDE attractor}.~This allows us to analyze the late- and early-time cosmology independently by examining the phase space of each  exponential separately.~The fixed points for a single exponential are given in Table~\ref{tab:fixed_points_table}.~

Beginning with early times when $\lambda_*=\alpha$, during the radiation epoch we can identify point G in Table~\ref{tab:fixed_points_table} with EDE supplying a fractional energy density\footnote{Technically, the expression in equation \eqref{eq:EDE_saddle_density} corresponds to point B* in Table~\ref{tab:general_fixed_points}.~Point G in Table~\ref{tab:fixed_points_table} is found by subsituting $w=1/3$ corresponding to the radiation era.~We keep equation \eqref{eq:EDE_saddle_density} more general for the purposes of a later discussion.} 
\begin{equation}
\Omega_\phi=\frac{3(1+w)}{\alpha^2} \simeq f_{\rm ede} \,.
\label{eq:EDE_saddle_density}
\end{equation} 

Previous EDE analyses suggest that the $H_0$ tension will be resolved if $f_{\rm ede}\sim 0.1$ \cite{Poulin:2023lkg}, implying that $\alpha\approx\sqrt{40}$.~This value of $\alpha$ is consistent with point $G$ being a saddle point.~At late times, when $\lambda_*=\beta$, the system will flow towards point $E$ or $F$ depending on the value of $\beta$.~For $\beta<\sqrt{3}$, the global attractor is point $E$, which corresponds to a scalar-dominated universe with equation of state 
\begin{equation}
    w_{\phi}=-1+\frac{\beta^2}{3} \,.
    \label{eq:EoS}
\end{equation}
Given that current observations indicate that the equation of state (EoS) of dark energy is $w\approx-1$ to high accuracy \cite{DES:2021wwk}, we expect small values of $\beta\ll1$ to correspond to LDE.~The choices above imply that point F corresponds to a saddle during the matter epoch where the scalar accounts for a fraction of the matter density;~we refer to this point as the \textit{dark matter saddle}.~

\begin{figure}[h]
    \centering
    \includegraphics[width=3.33in]{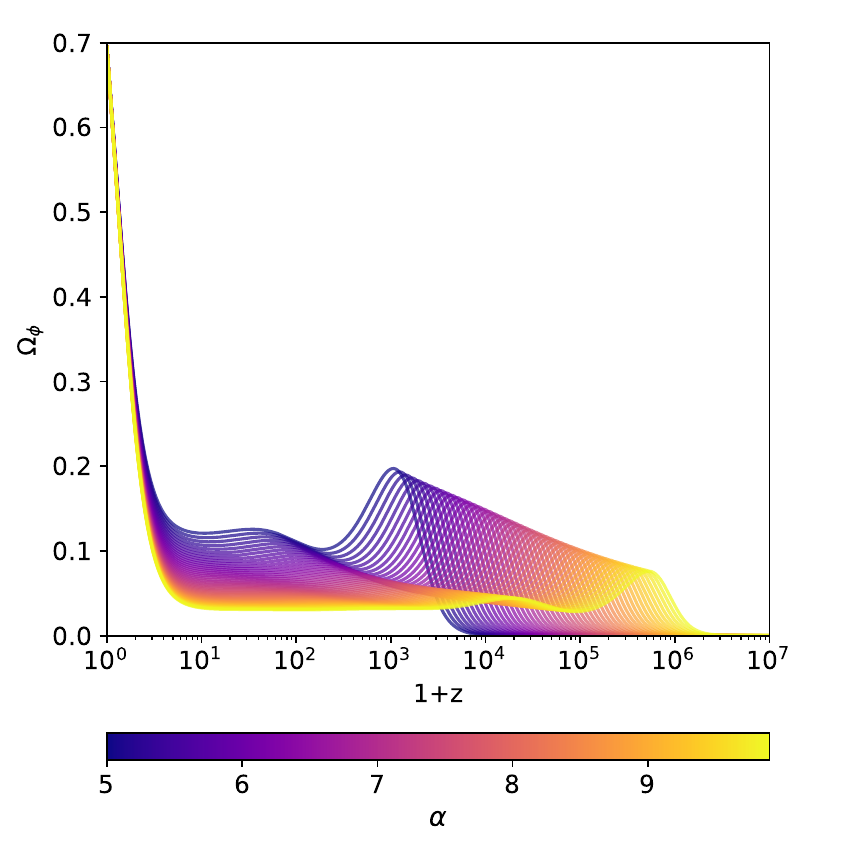}
    \caption{
    Background evolution of $\Omega_\phi$ as a function of redshift when $\alpha$ is varied.~We fixed $\beta= 0.01$,  $\log_{10}V_\alpha=-7.8$.~The \LCDM\ parameters were fixed to the \LCDM\ best fit given in Table~\ref{tab:marginalized}.~Increasing $\alpha$ reduces the fractional EDE  injection and shifts this injection to earlier times.~
    }
    \label{fig:alpha}
\end{figure}

\begin{figure}[h]
    \centering
\includegraphics[width=3.33in]{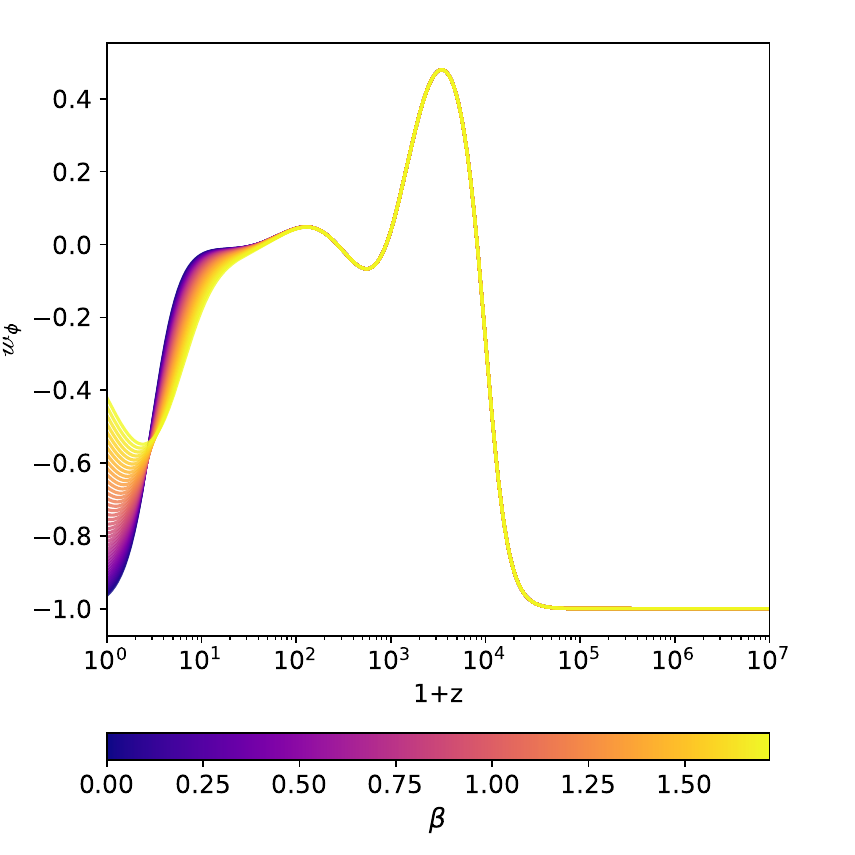}
    \caption{
    Evolution of the EoS $w_\phi$ of $\phi$ as a function of redshift when varying $\beta$.~Same as Fig.~\ref{fig:alpha}, we fixed $\alpha= \sqrt{40}$ and $\log_{10}V_\alpha=-7.8$.~The case $\beta=0$ corresponds to a cosmological constant driving LDE.~Note that $\beta$ only becomes relevant in the LDE era around $z\sim O(10)$.~
    }
    \label{fig:beta}
\end{figure}

\begin{figure}[ht]
    \centering
    \includegraphics[width=3.33in]{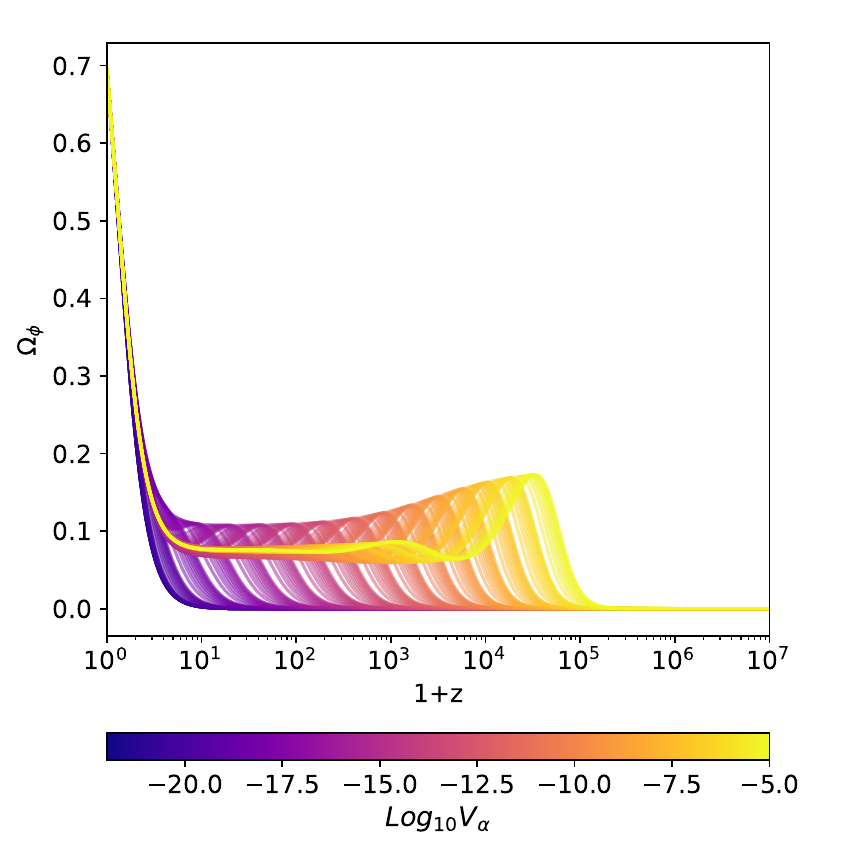}
    \caption{
    The evolution of $\Omega_\phi$ as a function of redshift when varying $V_{\alpha}$.~Here, we fixed $\alpha= \sqrt{40}$ and $\beta=0.01$.~The \LCDM\ parameters were fixed as in Fig.~\ref{fig:alpha}.~Increasing $V_{\alpha}$ leads to a larger and earlier injection.~Although increasing $\alpha$ has a similar effect of shifting the injection to higher redshifts, it decreases the amplitude of injection.    }
    \label{fig:v_alpha}
\end{figure}

We can verify the theoretical predictions above via a qualitative exploration of the  parameter space.~This exploration is also helpful for informing our subsequent data analysis.~Figs.~\ref{fig:alpha}, \ref{fig:beta}, and \ref{fig:v_alpha} show the effects of varying $\alpha$, $\beta$, and $V_\alpha$.~We only show the most impacted quantity for each parameter variation:~$\alpha$  and $V_\alpha$ predominantly effect $\Omega_\phi$ while $\beta$ sets the equation of state of LDE.

Beginning with Fig.~\ref{fig:alpha}, one can see that larger values of $\alpha$ reduce the amplitude of the EDE injection, consistent with the predictions of equation \eqref{eq:EDE_saddle_density}.~As for the injection redshift $z_c$, varying $\alpha$ changes the effective mass $m$ of the field given by 
\begin{equation}
m(\phi)=\alpha\frac{\sqrt{V_\alpha}}{\mpl} e^{- \frac{\alpha\phi}{2\mpl}} \,,
    \label{eq:mass}
\end{equation}
and the injection occurs when $m(\phi(z))=H(z)$.~For injections in the radiation era, $z_c$ can be approximated as
\begin{equation}
    z_c \sim \left(\frac{V_\alpha \alpha^2 e^{-\alpha\frac{\phi}{\mpl}}}{\Omega_r H_0^2\mpl^2}\right)^{1/4}-1.
    \label{eq:z_c}
\end{equation}
Hence, as $\alpha$ increases, the injection is pushed to higher redshifts, as can be seen in Fig.~\ref{fig:alpha}.~ 

Turning to Fig.~\ref{fig:beta}, which demonstrates the effect of varying $\beta$, in accordance with our predictions above there is no effect at high redshifts but the equation of state of dark energy is increased for larger values of $\beta$ as per equation \eqref{eq:EoS}.

Finally, Fig.~\ref{fig:v_alpha} reveals that $V_{\alpha}$ exhibits significant degeneracy with $\alpha$.~A larger $V_\alpha$ increases the field's mass according to equation \eqref{eq:mass} and directly shifts the injection towards higher redshifts.~It also changes the  amplitude of the energy injection indirectly because this depends on the background EoS $w(z)$ as shown in equation \eqref{eq:EDE_saddle_density}.~Since $w$ increases with redshift, so does the injection amplitude.

In light of the above, it is possible that some amount of tuning of the parameters/initial conditions may be needed to resolve the $H_0$ tension, although  less than canonical EDE models.\footnote{The tuning in $\beta$ is the usual tuning associated with quintessence models, and is not a new feature of @EDE.}~Whether or not this is tantamount to a fine-tuning depends upon the nature of the potential and whether these parameter choices are radiatively-stable.~In the case of our fiducial model, the double exponential arises in string theory \cite{Barreiro:1999zs}, so one expects the free parameters to be fixed in the UV.~The question is then one of finding string theory realizations that fix the parameters to values that can simultaneously resolve the $H_0$ tension, and explain LDE.~

\begin{figure}
    \centering
    \includegraphics[width=0.45\textwidth]{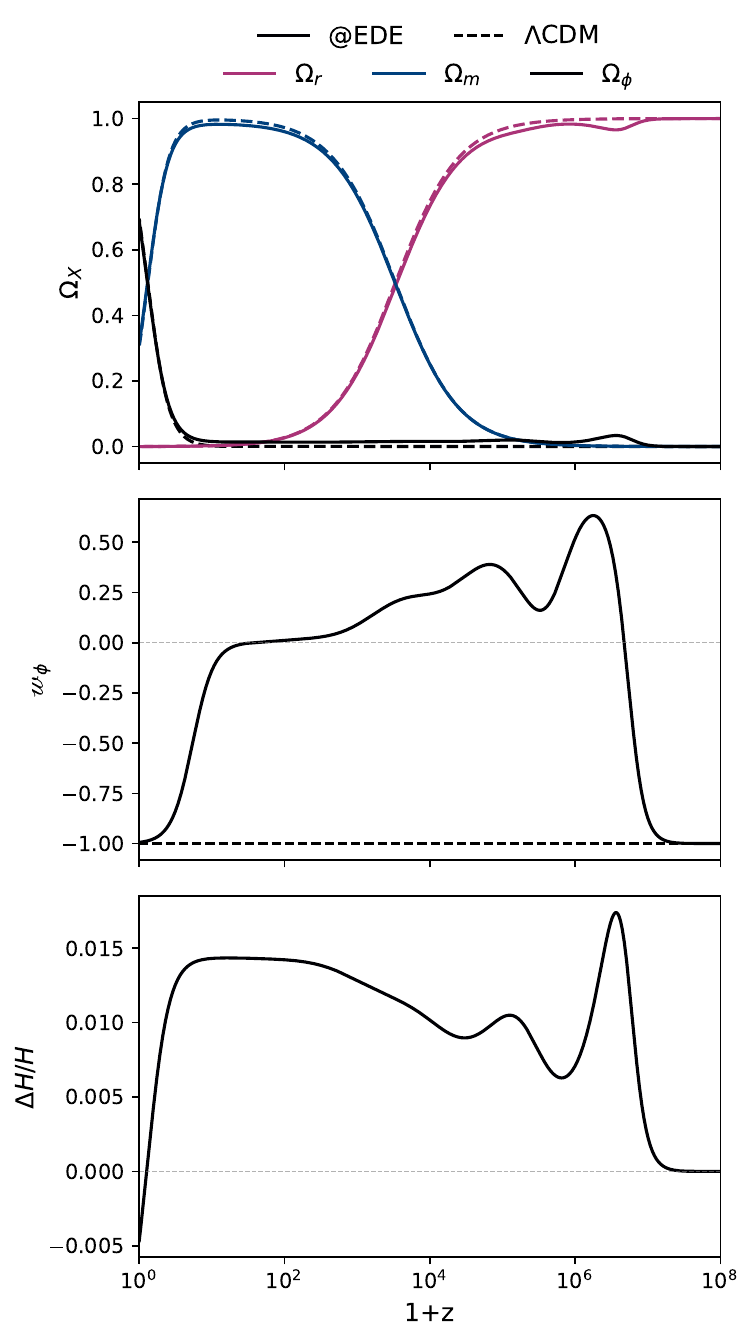}
    \caption{
    Evolution of important cosmological parameters in @EDE (solid curves) relative to the \LCDM\ best fit (dashed).~The @EDE cosmology shown is the best-fit from the \textit{narrow priors} exploration in Sec.~\ref{Section IV.A Narrow @EDE} that excludes \LCDM\ from the prior, with best fits given in Table~\ref{tab:marginalized}.~When \LCDM\ is included in the allowed @EDE parameter space, the resultant best fit is indistinguishable from \LCDM.~\textit{Top}:~Fractional energy densities $\Omega_X$ in matter, radiation, and the @EDE scalar relative to the total energy density of the Universe.~\textit{Middle}:~The equation of state $w_\phi$ of @EDE, which begins frozen with $w_\phi = -1$, becomes dynamical around $z\sim10^5$ acting as EDE, redshifts like matter during the matter era with $w_\phi \simeq 0$, and finally ends with scalar-field domination with $w_\phi$ close to -1 today.~\textit{Bottom}:~The fractional change in $H(z)$ induced by @EDE relative to \LCDM.~
    }\label{fig:background_plots}
\end{figure}

To summarize our scenario, we expect that the universe will approach an EDE saddle-point in the radiation epoch, transition to a matter saddle with some fraction of dark matter composed of the (non-clustering) scalar, and will ultimately settle into the LDE attractor.~This progression is shown in Fig.~\ref{fig:background_plots}, where we plot the evolution of the fractional energy densities  in \LCDM\ and @EDE, the equation of state $w_\phi$ of @EDE, and the modification to the expansion rate $\Delta H/H$.~The introduction of @EDE increases $H(z)$ at early times with a localised injection, then settles into a constant fractional increase, ending with $H(z)$ smaller than in \LCDM\ because the @EDE equation of state $w_\phi > -1$.~Note that \LCDM\ parameters are not fixed across cosmologies in Fig.~\ref{fig:background_plots} because such cosmologies are already shown by Figs.~\ref{fig:alpha}-\ref{fig:v_alpha}.

We now confront our fiducial @EDE model with data.~Ultimately, we find that the scenario described above is excluded with high significance for reasons we delineate below.~We use this insight to discuss how future @EDE model-building efforts may be able to construct theories that better-fit the data.~

\section{Methodology and Data}
\label{Section III:Methodology}

We modified the Cosmic Linear Anisotropy Solving System (\texttt{CLASS}) \cite{Blas:2011rf,Lesgourgues:2011re} to 
numerically solve the coupled Friedmann-Klein-Gordon equations of the double-exponential potential and evolve its perturbations.~To sample the @EDE and \LCDM\ parameter space and capture any multimodality we used PolyChord (\texttt{PC}) \cite{Handley:2015fda} --- 
which implements the nested-sampling algorithm \cite{Skilling:2006gxv} --- wrapped by the sampler \texttt{Cobaya} \cite{Torrado:2020dgo}.~We analyzed our results using GetDist \cite{Lewis:2019xzd}.~
For \texttt{PC}, we set the number of live points to $n_{\rm live}=35D$ where $D$ is the dimensionality of the parameter space and use the default convergence criteria  $\mathcal{Z}_{\rm live}= 10^{-2} \mathcal{Z} $ where $\mathcal{Z}_{\rm live}$ is the posterior mass contained in the live points and $\mathcal{Z}$ is the total model evidence.~

\subsection{Datasets}
\label{Section III.A :Datasets}
We used the data combination that has  become standard for EDE and Hubble tension investigations \cite{Poulin:2023lkg} in order to ensure a like-for-like comparison with previous works:~
 
\begin{enumerate}
    \item  \textbf{Cosmic microwave background (CMB) data}:~
    We used CMB data from Planck 2018 \cite{Planck:2018vyg} comprising of the low-$\ell$ TT and EE data, along with the high-$\ell$ \texttt{plik-lite} likelihood  and the gravitational lensing data \cite{Planck:2018lbu}.~    We used \texttt{plik-lite} instead of the full \texttt{plik} likelihoods to reduce the dimensionality of the parameter space, removing nuisance parameters that would significantly slow down \texttt{PC}. The datasets consist of $N_{CMB}=2352$ data points.~
    
    \item \textbf{Baryon acoustic oscillations (BAO)}:~We used BAO data that is consistent with the CMB including the 6dF galaxy survey (6dFGS) \cite{Beutler:2011hx}, SDSS DR7 \cite{Ross:2014qpa} and, SDSS-III DR12 \cite{BOSS:2016wmc} which, combined, constrain the distance-redshift relation with $N_{BAO}=5$ data point spanning  $0.01<z<0.61$ using spectroscopic methods on galactic data.
  
    \item \textbf{Supernovae (SNe)}:~We used the Pantheon \cite{Pan-STARRS1:2017jku} data which surveyed the redshifts and magnitudes of 1048 Type Ia Supernovae ranging from $0.01<z<2.3$.~An updated version, Pantheon+ was released while this manuscript was in preparation.~We do not expect our results to change with updated data as the CMB strongly constrains and excludes the fiducial @EDE scenario explored here.~
    
    \item \textbf{Cepheid distance ladder $H_0$}:~We use the SH0ES direct measurement of $H_0=\left(74.03\pm1.42\right)$km/s/Mpc \cite{Riess:2019cxk} as a best-case test scenario to check whether our scenario resolves the Hubble tension and to mitigate prior volume projection effects in EDE cosmologies \cite{Smith:2020rxx, Herold:2021ksg}.~A new, more precise, measurement \cite{Murakami:2023xuy} of $H_0=\left(73.29\pm0.90\right)$ km/s/Mpc was released as we prepared this manuscript, but updating the $H_0$ likelihood we use would have a minimal impact on our results because CMB data dominate our constraints.
\end{enumerate}

\subsection{Parameter space}
\label{Section III.B Parameter}

To explore @EDE cosmology, we varied the standard \LCDM\ parameters:~the physical densities of baryons $\omega_b$ and cold dark matter $\omega_c$, the amplitude $A_s$ of the primordial power spectrum as $\ln 10^{10} A_s$ and its tilt $n_s$, the optical depth $\tau$ due to reionization, and the expansion rate $H_0$ of the Universe today.~We additionally varied the @EDE parameters $\alpha$, $\beta$, and $V_\alpha$ that control the @EDE scalar potential.~We fixed the remaining @EDE parameters as follows.

We fixed the initial condition $\phi_i$ by exploiting a symmetry of the potential.~Specifically, the action is invariant under $\phi\rightarrow\phi+\phi_0$, $V_k\rightarrow V_ke^{k\frac{\phi_0}{\mpl}}$ with $k=\alpha,\,\beta$ and where $\phi_0$ is a constant.~This allows us to fix $\phi_i$ without loss of generality.~We make the arbitrary choice to fix $\phi_i = -4.583 \mpl$.~

We fixed $\dot{\phi}_i$ using attractor initial conditions.~Since the field starts frozen in time, we determined an attractor initial condition for $\dot{\phi_i}$ in \texttt{CLASS} by setting $\Ddot{\phi} = 0$ in the equation of motion \eqref{eq:klein-gordon} i.e., we assumed that the field is only slowly-rolling, and approximated the potential as $V(\phi) \approx V_\alpha e^{-\alpha\frac{\phi}{\mpl}}$ since the $\beta$-exponential is only relevant at late times.~With these approximations, we then have $\dot{\phi_i} = \frac{\alpha V_\alpha}{3H_i}e^{-\alpha\frac{\phi_i}{\mpl}}$ as our initial condition.~

Finally, at late times, the system is determined by the $\beta$-exponential and approaches fixed point E in Table~\ref{tab:fixed_points_table} with $\lambda_*=\beta$.~The scalar acts as LDE with an EoS given by equation \eqref{eq:EoS}.~To close the universe, we therefore modified \texttt{CLASS} to shoot for $V_\beta$ using $V_\beta=3 H_0^2 \mpl^2\Omega_\phi$ for a given set of other parameters.~Fixing $\phi_i$, $\dot{\phi}_i$, and $V_\beta$ reduced the dimensionality of the @EDE cosmology parameter space from 12 dimensions to 9 dimensions (the six $\Lambda$CDM parameters along with $\alpha$, $\beta$, and $V_\alpha$).~

\begin{table}[h]
    \begin{tabular}{|\Col|\Col|\Col|}
        \hline
        Parameter & Wide Prior & Narrow Prior\\
        \hline \hline
         $\alpha$ & $\left[0,17\right]$ & $\left[8.5, 15\right]$ \\
         $\log_{10}V_\alpha$ & $\left[-50,-6\right]$ & $\left[-15, -9\right]$\\
         $\beta$ & $\left[0,\sqrt{3}\right]$ & $\left[0, \sqrt{3}\right]$ \\
         \hline
         $f_{ede}$ & $\left[0,0.268\right]$ & $\left[0.033, 0.100\right]$\\
         $\log_{10}z_c$ & $\left[3,9.82\right]$ & $\left[3.13, 8.06\right]$\\
        \hline
    \end{tabular}
    \caption{
    Priors for @EDE parameters that include \LCDM\ (wide priors) and exclude it (narrow priors).
    }  
    \label{tab:priors}
\end{table}

We used uninformative priors for the \LCDM\ parameters and the priors given in Table~\ref{tab:priors} for the @EDE parameters.~The \textit{wide @EDE prior} range explores $0 \leq f_{\rm ede} \leq 0.15$ and $z_c \geq 1100$, in terms of the usual EDE parameters:~the maximal fractional energy density $f_{\rm ede}$ in EDE that occurs at redshift $z_c$.~Ultimately, we found that the best-fitting @EDE model is nearly identical to $\Lambda$CDM, and that our expected scenario described in the previous section is excluded with high significance.~To help to understand why this is the case, we performed a second analysis with \textit{narrow priors} centered on the theoretical values derived in section (\ref{Section II.B:Background dynamics}).~This restricted prior range focuses on $10^3 \leq z_c \leq 10^8$ and excludes \LCDM\ by forcing an @EDE energy injection of $0.033 \leq f_{\rm ede} \leq 0.1$.~These cosmologies, along with \LCDM, are explored in the following section.~

\section{Results}\label{Section IV:Results}

We begin by exploring an @EDE scenario with wide priors that includes \LCDM\ as a nested model.~In this prior range, \LCDM\ is recovered when $\alpha=0$ and $V_\alpha=0$ such that there is no EDE phase, and $\beta = 0$, which implies a cosmological-constant LDE.~This allows for a direct comparison of the goodness-of-fits, constraints on $H_0$, and any preference for @EDE over \LCDM.~This model is labelled \textit{@EDE:~wide priors} in the following.~

\begin{figure*}
\centering
\includegraphics[width=7.05in]{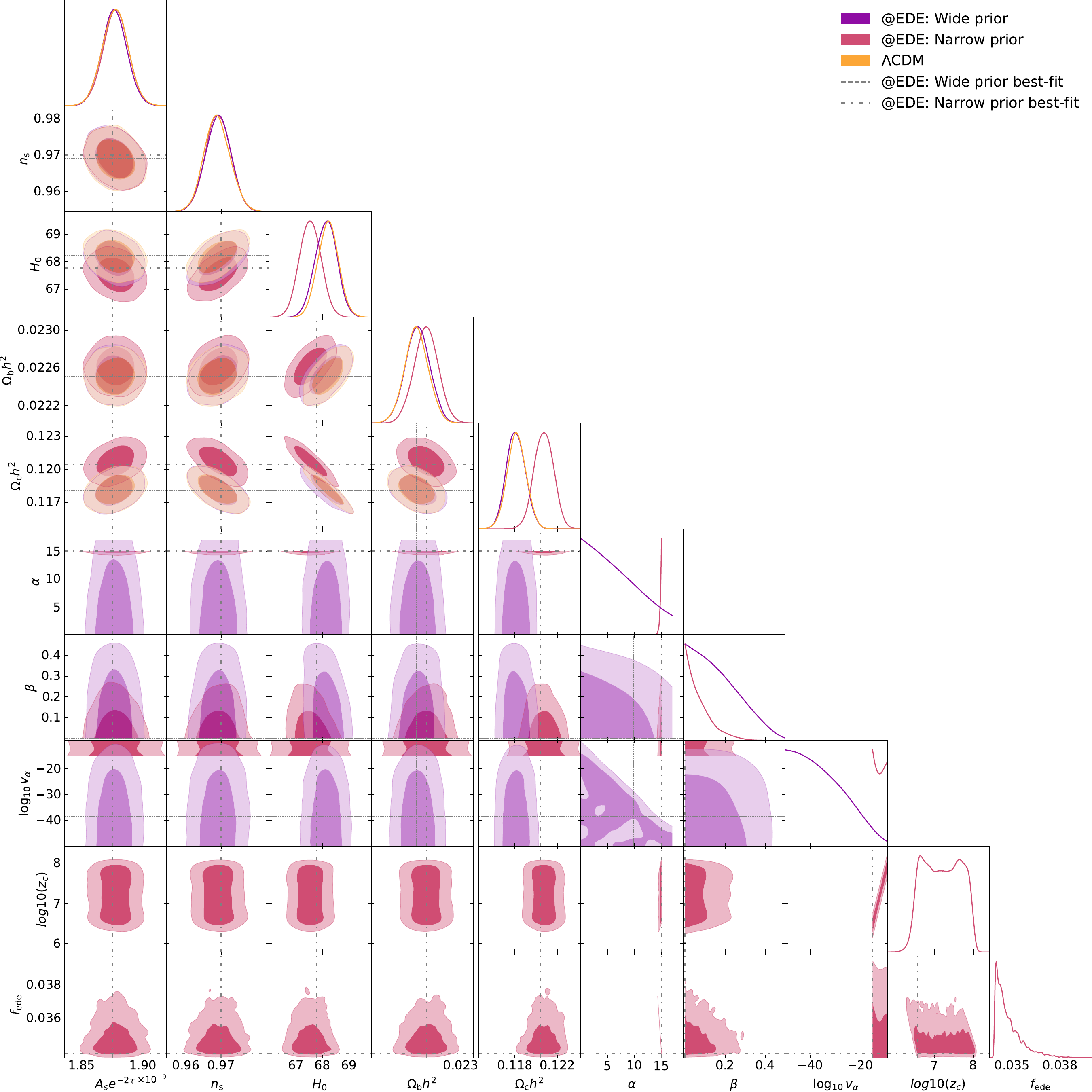}
    \caption{Posteriors for various cosmologies when fit to Planck2018+BAO+SNe+SH0ES.~We explore @EDE with wide priors that include \LCDM\ in purple, @EDE with narrow priors that exclude \LCDM\ in red, and \LCDM\ in orange.~The horizontal and vertical lines mark the best-fit points of the wide prior (densely dashed) and narrow prior (dash-dotted) @EDE cosmologies.~
    }
\label{fig:contours}
\end{figure*}

Despite theoretical expectations, we find that @EDE is not preferred over \LCDM, and therefore does not offer a resolution to the Hubble tension.~Data prefer a scenario that closely, within $1\sigma$, mimics \LCDM, effectively excluding any early injection of energy density that may resemble EDEs, as shown in Fig.~\ref{fig:contours}.~Effectively, @EDE remains frozen throughout cosmic history, acting like a cosmological constant with $w_\phi = -1$ and with the scalar potential dominated by $V_\beta e^{-\beta \frac{\phi}{\mpl}}$.~This cosmology is indistinguishable from \LCDM\ in the top and bottom panels of Fig.~\ref{fig:background_plots}.~
Its best-fitting parameters are given in Table~\ref{tab:marginalized}.

This strong preference for \LCDM\ is not due to prior volume effects of the model as can be seen from Table~\ref{tab:marginalized}, wherein the best-fits are well within $1\sigma$ of the means, demonstrating that the peak of the likelihood coincides well with the peak of the samples at the mean.~As @EDE posteriors are completely consistent with \LCDM\ posteriors, it follows that our fiducial @EDE model does not resolve the Hubble tension, with an insufficient increase of $\Delta H_0 = 0.15$ km/s/Mpc.~Moreover, despite three additional parameters in @EDE, we find a negligible improvement in goodness-of-fit with $\Delta \chi^2=-0.28$.~The expectation for three additional parameters would be a minimal improvement of $\Delta \chi^2 = 3$, which further underscores the data preference for \LCDM\ over @EDE.~

To further understand the shortcomings of our fiducial @EDE model, and to gain insight into building a  model that can better fit the data, resolve the Hubble tension, and link early and late dark energy, we next explore forcing an @EDE injection through a prior-restricted @EDE model that excludes \LCDM, labeled as \textit{@EDE:~narrow priors} in what follows.

\def\arraystretch{1.6}
\begin{table*}[t]
\begin{tabular}{|\Col|\Col|\Col|\Col|}\hline
        Parameter & \LCDM\ & @EDE:~wide priors & @EDE:~narrow priors \\ 
        \hline\hline
        $\log(10^{10} A_\mathrm{s})$  & $3.052(3.045)\pm 0.015$ & $3.051(3.052)^{+0.013}_{-0.016}$ & $3.083(3.085)\pm 0.017$ \\ 
        \hline
        $n_\mathrm{s}$ & $0.9689(0.9687)^{+0.0034}_{-0.0039}$ & $0.9691(0.9691)\pm 0.0036$ & $0.9691(0.9700)\pm 0.0036$ \\ 
        \hline
        $\omega_\mathrm{b}$ & $0.02251(0.02249)\pm 0.00013$ & $0.02253(0.02251)\pm 0.00013$ & $0.02262(0.02262)\pm 0.00013$ \\ 
        \hline
        $\omega_\mathrm{c}$ & $0.11816(0.11840)\pm 0.00088$ & $0.11808(0.11807)\pm 0.00089$ & $0.12072(0.12043)\pm 0.00091$ \\ 
        \hline
        $\tau_\mathrm{reio}$ & $0.0599(0.0593)\pm 0.0076$ & $0.0597(0.0599)^{+0.0066}_{-0.0080}$ & $0.0751(0.0769)\pm 0.0088$ \\ 
        \hline
        ${D_A}$ & $(12.76)$ & $(12.76)$ & $(12.62)$ \\
        \hline
        $r_{s*}$ & $(144.85)$ & $(144.92)$ & $(143.12)$ \\
        \hline
        $100\theta_*$ & $(1.042009)$ & $(1.042049)$ & $(1.040470)$ \\
        \hline
        $H_0$ & $68.21(68.09)\pm 0.39$ & $68.13(68.24)\pm 0.41$ & $67.53(67.77)\pm 0.41$ \\
        \hline
        $\sigma_8$ & $0.8084(0.8084)\pm 0.0061$ & $0.8069(0.8080)^{+0.0058}_{-0.0066}$ & $0.7729(0.7758)\pm 0.0069$ \\
        \hline
        $S_8$ & $0.813(0.815)\pm 0.010$ & $0.813(0.812)\pm 0.010$ & $793(0.792)\pm 0.010$ \\
        \hline\hline
        $\alpha$ & - & $< 8.53(9.82)$ & $> 14.8(15)$ \\ 
         \hline
        $\beta$ & - & $< 0.213(0)$ & $< 0.0883(1.69\times 10^{-6})$ \\ 
        \hline
        $\log_{10}V_{\alpha}$ & - & $< -30.9(-38.5)$ & $(-15)$ \\
        \hline
        $f_{\rm ede}$ & - & - & $0.03480(0.03388)^{+0.00017}_{-0.00092}$ \\ 
        \hline
        $\log_{10}z_c$ & - & - & $7.20(6.56)\pm 0.46 $ \\ 
        \hline\hline
        %%%%%% chi2 %%%%%
        % CMB
        {$\chi^2_{CMB} (\Delta)$} & $1013.99$ & $1014.64(+0.65)$ & $1052.70(+38.71)$ \\
        %BAO
        {$\chi^2_{BAO} (\Delta)$} & $5.23 $ & $5.24(+0.01)$ & $7.98(+2.75)$ \\
        {$\chi^2_{H_0}(\Delta)$} & $15.45$ & $14.56(-0.89)$ & $17.45(+2.00)$ \\
        % \hline
        {$\chi^2_{Pantheon}(\Delta)$} & $1034.82$ & $1034.77(-0.05)$ & $1035.41(+0.59)$ \\
        \hline
        \hline
        {$\chi^2(\Delta)$} & $2069.49$ & $2069.21(-0.28)$ & $2113.54(+44.05)$ \\
        {$\chi^2_{red}(\Delta)$} & $0.6089$ & $0.6093(+0.0004)$ & $0.6224(+0.0135)$ \\
        \hline
    \end{tabular}
    \caption{
    Marginalized posteriors for \LCDM\ and two @EDE cosmologies with different priors, showing mean (best-fit) $\pm 1\sigma$.~
    The @EDE scenario with wide priors includes \LCDM\ as a nested model, while the narrow priors exclude \LCDM.~
    The wide-priors model does not list $f_{\rm ede}$ and $z_c$ as there is no EDE phase in that scenario, just LDE.~
    We also show the various $\chi^2$ $(\Delta \chi^2)$ broken down by data set and their differences relative to \LCDM.~
    }
  \label{tab:marginalized}
\end{table*}

\subsection{Excluding \LCDM\ from the parameter space}\label{Section IV.A Narrow @EDE}

The narrow priors constrain $0.033 \leq f_{\rm ede} \leq 0.1$ and $10^3 < z_c \leq 10^8$, forcing a minimum @EDE energy injection of $3.3\%$.~Note however that these phenomenological parameters are not explored via uniform distributions.~We employ flat priors on the model parameters $\alpha$ and $V_\alpha$, such that the smallest injection occurs at $z_c \simeq 3.6 \times 10^6$, while the largest $f_{\rm ede}$ occurs at $z_c \simeq 5 \times 10^4$.~

In this restricted scenario, the data prefer $\alpha=15$ and $\log_{10} V_\alpha=-15$, both at the edges of their respective narrow-prior ranges as seen from comparing with Table~\ref{tab:priors}, resulting in $z_c \simeq 3.7 \times 10^6$ and $f_{\rm ede} \simeq 3.4\% $.~This best fit reduces the impact of @EDE on observables via minimizing the energy injection and consequently, pushes @EDE dynamics into the redshift range that data are less sensitive to.~The Planck CMB is sensitive to new physics injections up to a maximum redshift of $z \lesssim 10^6$ \cite{Karwal:2016vyq, Berghaus:2022cwf,Linder:2010wp}.~For earlier injections, @EDE effectively acts like an additional matter-tracking component, similar to tracking early dark energies \cite{Calabrese:2011hg, Pettorino:2013ia, Doran:2006kp} and will map onto the same upper-limit constraints as tracker early dark energies.~

As we force a non-zero @EDE energy injection, \LCDM\ parameters are forced to compensate for its impact, inducing the parameter shifts shown in Fig.~\ref{fig:contours} and Table~\ref{tab:marginalized}.~These shifts can be understood within the context of the CMB as follows.~First, at $z \sim 10^7$, the scalar field begins to roll and enters an EDE phase, contributing $f_{\rm ede} \simeq 3.39\%$ to the energy budget of the Universe.~Usually, after this early-universe peak in $\Omega_\phi$, EDEs dilute away and have no further impact on cosmology.~However, @EDE loses roughly half its fractional energy density, then approaches the matter attractor, diluting like matter and contributing $1.33\% < \Omega_\phi <  1.37\%$ to the energy budget of the Universe.~This contribution persists until the dark energy attractor takes over and @EDE contributes the LDE that dominates the Universe today.~

If the $\Lambda$CDM parameters are fixed, the addition of @EDE decreases the size of the sound horizon $r_s$. 
The existence of this additional component in the early universe that does not cluster like matter makes the gravitational potentials $\Phi + \Psi$ in the Newtonian gauge \cite{Ma:1995ey} decay relative to a $\Lambda$CDM universe, enhancing the early integrated Sachs-Wolfe (ISW) effect. 
It also increases $H(z)$ at early times, making modes enter the horizon sooner and shifting the oscillations in the Sachs-Wolfe (SW) effect at recombination. 
In addition, the lingering component of @EDE in the late universe provides an additional contribution to the late ISW effect, and reduces the angular diameter distance $D_A$ to the CMB.
The decrease in $D_A$ cannot compensate the decrease in $r_s$ and the precisely measured angular size $\theta_*$ of the sound horizon will decrease.
 Altogether, @EDE boosts power in the first CMB peak and lower multipoles, as well as shifts the CMB peaks.

When the \LCDM\ parameters are allowed to vary, they react to this additional component and compensate for the effects above through shifts in $H_0$, $\omega_b$ and $\omega_c$ as follows.~A substantial increase in $\omega_c$ and slight increase in $\omega_b$ suppresses the TT power and counteracts the enhanced early and late ISW effects  as well as the shifts in the SW effect at recombination.  
As the densities of both matter species increase, matter-radiation equality occurs sooner and at higher $z_{\rm eq}$, which counters the decays in $\Phi + \Psi$. 
The increase in $\omega_c$ also impacts both $r_s$ and $D_A$, such that $\theta_*$ would increase to larger than the observed value.~The increase in $\omega_b$ and slight decrease in $H_0$ then try to offset this shift in $\theta_*$ and the CMB peak locations.

\begin{figure*}[t]
\centering
\includegraphics[width=0.49\textwidth]{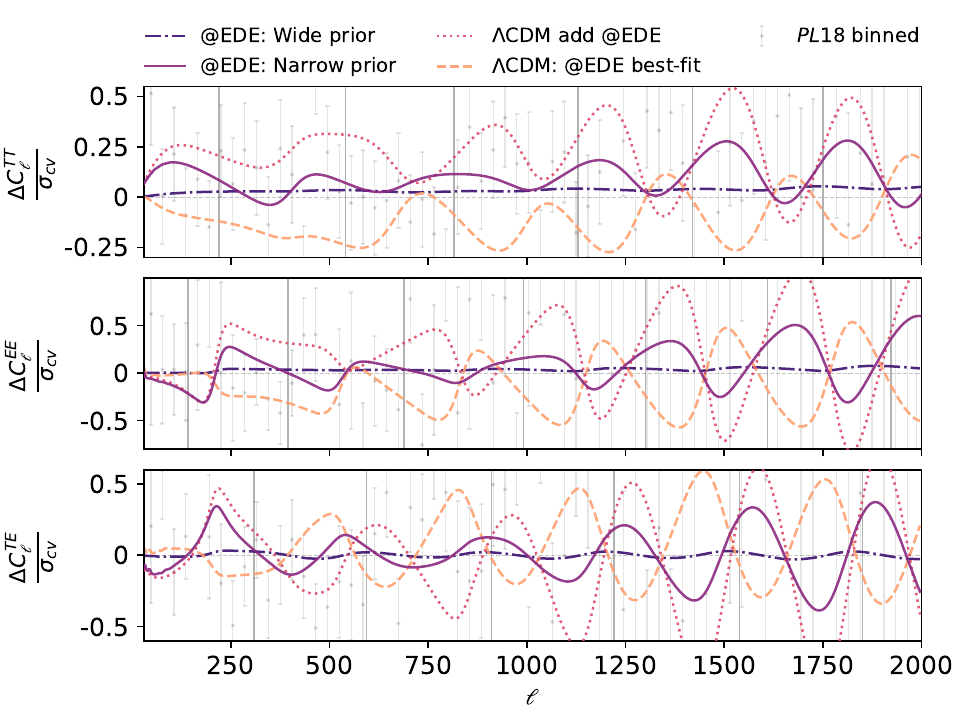}
\includegraphics[width=0.49\textwidth]{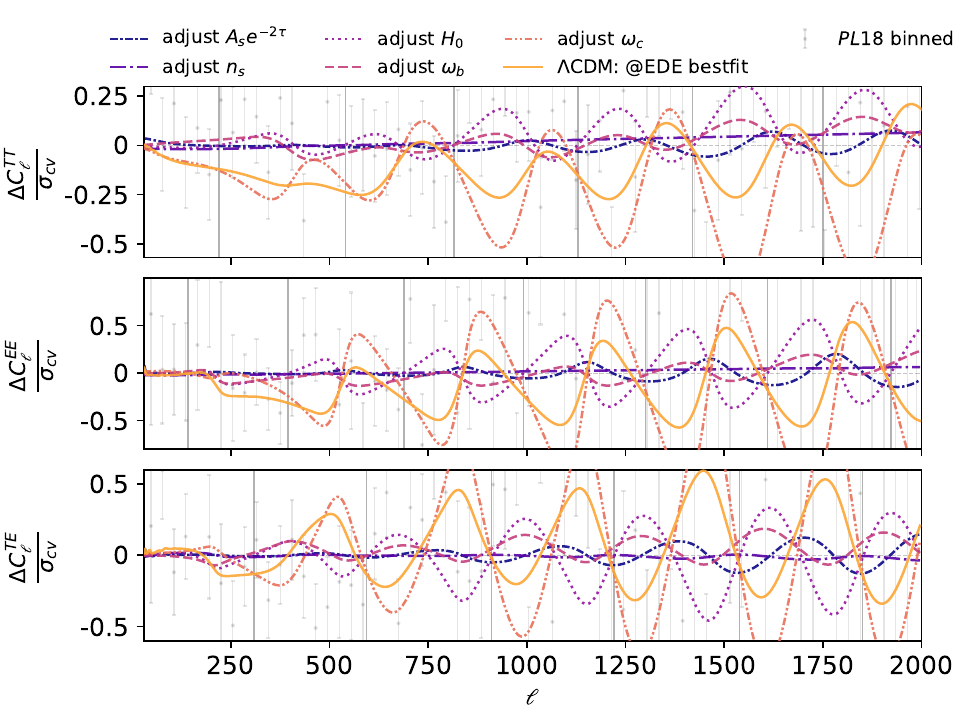}
    \caption{
    CMB residuals for various cosmologies relative to the \LCDM\ best fit as a fraction of cosmic variance.~The light gray data points are binned measurements from Planck 2018.~The solid dark grey vertical lines mark peak locations in all spectra.~\textit{Left:} The dash-dotted blue curve shows the @EDE best-fit for wide priors that include \LCDM, while the solid purple curve excludes \LCDM\ with narrow priors for @EDE.~The dashed orange curve is a \LCDM\ cosmology but with parameters set to the best fit found for the @EDE narrow prior scenario.~The dotted red curve on the other hand is an @EDE cosmology, with \LCDM\ parameters set to the \LCDM\ best fit and @EDE parameters at the same narrow-prior best fit.~With these two curves, we show how \LCDM\ and @EDE parameters trade off, and how they compensate for each other or, rather, fail to do so in an attempt to better fit data.~Effectively, the dashed orange and dotted red residual curves sum up to the solid purple.~\textit{Right:} We further break down the dashed orange curve on the left figure (shown here in solid orange) into its component shifts, shifting one \LCDM\ parameter at a time away from its \LCDM\ best fit.~The greatest shifts are due to $\omega_c$ (an overall suppression of the power spectrum and a shift in the peaks to larger angular scales) and $H_0$ (a compensating shift in the peaks to smaller angular scales).    }
\label{fig:CMB_residuals}
\end{figure*}

The right panel of Fig.~\ref{fig:CMB_residuals} shows these competing effects as the \LCDM\ parameters adapt to absorb the impact of @EDE.~These parameter shifts are ultimately unsuccessful, leading to the significantly worsened fit to CMB data with $\Delta \chi^2 = +44.05$ relative to \LCDM.~The fit to BAO data is also worsened with $\Delta \chi^2 = + 2.75$.~Observations of BAO are in excellent agreement with CMB data interpreted within the \LCDM\ paradigm, with agreement between $r_s$, $D_A$ and $H(z)$ at very different $z$ \cite{BOSS:2016wmc}.~The introduction of @EDE spoils this consistency and increases the BAO $\chi^2$.~

Ultimately, two competing effects keep @EDE from being a viable solution to the Hubble tension.~Fitting the height of the first CMB peak requires increasing $\omega_c$ to suppress the additional early ISW effect from the early-universe injection of @EDE \cite{Knox:2019rjx, Vagnozzi:2021gjh,Poulin:2023lkg}.~But fitting the location of the same peak requires decreasing $\omega_c$ to maintain a good fit to $\theta_*$ via shifts in $D_A$.~As both cannot be accommodated simultaneously, @EDE is not preferred by data and posteriors converge to a \LCDM-like universe.~

\section{Discussion and Conclusions}\label{Section V:Discussion}

In this paper we have proposed a unified framework for explaining early and late dark energy with a single scalar field --- attractive early dark energy.~This framework has the potential to simultaneously resolve the Hubble tension and drive the late-time acceleration of the cosmic expansion without the need to fine-tune the model parameters.~Instead, the coincidence between the onset of EDE and matter-radiation equality is explained by a saddle point of a dynamical system in the radiation epoch that attracts solutions independent of their initial conditions.~Similarly LDE arises naturally as a late-time global attractor of the system.~

We explored a fiducial @EDE model corresponding to a double exponential potential, which arises in string theory.~Unfortunately, this specific model was not preferred by the data because the dynamical system also possesses a saddle point during the matter era where the scalar contributes a fraction of non-clustering dark matter to the Universe's energy budget.~This reduces the angular diameter distance $D_A$ to the CMB.~In addition, as is the case with all EDE models, the early ISW effect is also enhanced.~The enhancement in the early ISW effect must be compensated by an increase in $\omega_c$, but the shift in $D_A$ must be compensated by a decrease in $\omega_c$.~Since, there is no single value of $\omega_c$ that can compensate both effects simultaneously, the model is a poor fit to the data.

It is prudent to discuss the potential for constructing viable @EDE models that can help to restore cosmological concordance.~Within the framework analyzed in the present work --- a four-dimensional phase space for the dynamical system --- we identified a second scalar potential that has the qualitative features required for the scalar to function as @EDE.~The near-identical background evolution and fixed points of this model imply that it is unlikely to be a better fit to the data.~We therefore discuss potential extensions of the framework.~One possibility  would be to consider potentials where the phase space is higher-dimensional.~This would enlarge the number of potential models, and introduce additional free parameters that would decouple the properties of the EDE and dark matter saddles.~One could then envision potentials where the amount of scalar dark matter during the matter epoch is negligible.~An alternative is to analyze non-minimal scalar theories that are known to form dynamical systems such as coupled quintessence \cite{Amendola:1999er}, where the scalar is conformally coupled to dark matter.~The fixed points of these theories are similar to those found in the uncoupled case but the equation of state for the scalar at each point depends on the strength of the dark matter coupling.~It may be possible to find models where the equation of state at the dark matter saddle is $w_\phi>0$ so that the scalar redshifts during the matter era, diminishing its contributions to the late universe which caused our fiducial model to be a poor fit to the data.~A similar effect could be achieved by studying K-essence theories, where the scalar has a non-canonical kinetic term (reference \cite{Bahamonde:2017ize} contains a comprehensive study of these models), or disformal quintessence \cite{Sakstein:2014aca,Sakstein:2015jca}, where the scalar is derivatively coupled to dark matter.~The derivative interactions induce a non-zero sound speed for the scalar, so that it would cluster on small scales, reducing the late ISW enhancement.

As remarked above,  the lack of a complete concordance model that is capable of explaining all of our observations is limiting our ability to interpret data from cosmological missions and astrophysical surveys.~This will remain the case as the current generation of missions conclude and the next generation begin to see first light.~Discovering the origin of the Hubble tension is therefore paramount.~If new physics underlies the tension then any fundamental description of said physics should be consistent with the principles of quantum field theory i.e., the model should be a natural effective field theory free of fine-tunings and coincidence problems.~In addition, it is natural that this new physics be connected with the other phenomena we observe in the universe e.g., late dark energy.~This work has taken steps in towards this by proposing a framework that ameliorates the fine-tuning and coincidence problems associated with early dark energy, and unifies both early and late dark energy into a single phenomenon driven by one new scalar degree of freedom.~Ultimately, our fiducial attractive early dark energy proposal did not provide a good fit to the data, but our study has provided novel lessons for future @EDE model-building efforts aimed at achieving this goal.

\begin{acknowledgments}
We are grateful for discussions with Eric Baxter, Jason Kumar, Vivian Poulin, David Rubin, and Istvan Szapudi.~The technical support and advanced computing resources from University of Hawai‘i Information Technology Services – Cyberinfrastructure, funded in part by the National Science Foundation CC\* awards \#2201428 and \#2232862 are gratefully acknowledged.
TK acknowledges support from NASA ATP Grant 80NSSC18K0694 and funds provided by the Center for Particle Cosmology at the University of Pennsylvania.~

\end{acknowledgments}

\bibliographystyle{apsrev4-1}
\bibliography{main}

\end{document}